\documentclass[12pt,a4paper]{article}
\usepackage{graphicx}
\usepackage{mylist}
\usepackage{latexsym,amssymb}
\usepackage[all]{xy}\CompileMatrices
\usepackage{color}
\usepackage{amsmath}
\definecolor{redex}{rgb}{.9,.1,.1}
\def\tr#1{\textcolor{redex}{#1}}
\def\al{\alpha}
\def\ba{\beta}
\def\ga{\gamma}
\def\Ga{\Gamma}
\let\da\delta

\def\la{\lambda} 

\def\ta{\theta}
\def\rar{\rightarrow}

\def\slar#1{\stackrel{#1}{\lrar}}
\def\lrar{\longrightarrow}

\def\Rar{\Rightarrow}

\def\mcl{\mathcal} 
\def\mbb{\mathbb} 
 
\def\mfr{\mathfrak}
\newtheorem{prop}{Proposition}\def\PRO{\begin{prop}}\def\ORP{\end{prop}}
\newtheorem{coro}{Corollary}\def\COR{\begin{coro}}\def\ROC{\end{coro}}
\newtheorem{theo}{Theorem}\def\TH{\begin{theo}}\def\HT{\end{theo}}
\newtheorem{defi}[prop]{Definition}\def\DE{\begin{defi}}\def\ED{\end{defi}}
\newtheorem{lemme}[prop]{Lemma}\def\LE{\begin{lemme}}\def\EL{\end{lemme}}
\newcommand{\AR}[2][c]{$$\begin{array}[#1]{lllllllllllllll}#2\end{array}$$}
\def\MA#1{\left(\begin{matrix}#1\end{matrix}\right)}
\def\EQ#1{\begin{eqnarray}#1\end{eqnarray}}
\def\mypar#1{\medskip\par\textbf{\emph{#1}}}
\def\emptyset{\varnothing}
\def\ens#1{\{#1\}}
\def\ie{\textit{i.e.}}

\def\st{^\star}

\def\qed{$\Box$}
\def\?{$\Box$}
\def\iN{\iota}
\def\oU{o}

\def\ei#1{e^{i#1}}
\def\emi#1{e^{{-i}#1}}
\def\eta{\ta}

\def\ad{\st}

\def\hil#1{\mfr H_{#1}}

\def\ket#1{{|}#1\rangle}
\def\bra#1{\langle#1{|}}
\def\oqb#1{\ket{\hskip-.1ex+_{#1}}}
\def\oqbn#1{\ket{\hskip-.1ex-_{#1}}}
\def\oqbb#1{\bra{\hskip-.1ex+_{#1}}}
\def\oqbnb#1{\bra{\hskip-.1ex-_{#1}}}

\def\ctR{\mathop{\wedge}\hskip-.4ex} 
\def\ctwo{{\mbb C}^2}  
\def\ztwo{{\mbb Z}_2}
\def\ost{\frac1{\sqrt2}}

\def\pit{\frac\pi2}
\def\norm#1{\|#1\|}
\def\brA#1{\rar_{#1}}
\def\Rr#1#2{J_{#1}^{#2}}
\def\Cx#1{\cx{#1}{}}
\def\Cz#1{\cz{#1}{}}
\def\cz#1#2{Z_{#1}^{#2}}
\def\cx#1#2{X_{#1}^{#2}}
\def\G{J}

\def\MS#1#2#3#4{{}^{#4}[{M}_{#2}^{#1}]^{#3}}
\def\ms#1#2#3{\MS{#1}{#2}{#3}{}}
\def\Ms#1#2{{M}_{#2}^{#1}}
\def\M#1#2{{M}_{#2}^{#1}}

\def\et#1#2{E_{#1#2}}
\def\CO#1{A_{#1}}
\def\ss#1#2{S_{#1}^{#2}}

\title{The Measurement Calculus}

\author{Vincent Danos\\Universit\'e Paris~7 \& CNRS\\
  {\small\texttt{Vincent.Danos@pps.jussieu.fr}}
  \and
  Elham Kashefi\thanks{This work was partially supported by the PREA, MITACS, ORDCF and CFI projects.}\\ IQC - University of Waterloo\\
  Christ Church - Oxford\\
  {\small\texttt{ekashefi@iqc.ca}}
  \and
  Prakash Panangaden\\ McGill University\\
  {\small\texttt{Prakash@cs.mcgill.ca}}}

\begin{document}
\maketitle
\begin{abstract}
We propose a calculus of local equations over one-way computing
patterns~\cite{mqqcs}, which preserves interpretations, and allows the
rewriting of any pattern to a standard form where entanglement is done
first, then measurements, then local corrections.  We infer from this that
patterns with no dependencies, or using only Pauli measurements, can only realise unitaries belonging to the Clifford group. 
\end{abstract}

\section{Introduction}
The \emph{one-way} model centres on 1-qubit measurements as the main
ingredient of quantum computation~\cite{mqqcs}, and is believed by
physicists to lend itself to easier implementations~\cite{Nielsen04,ND04,BR04,CMJ04}.
During computations, measurements and local corrections are allowed to
depend on the outcomes of previous measurements.

We first develop a notation for such classically correlated sequences of
entanglements, measurements, and local corrections.  Computations are
organised in patterns, and we give a careful treatment of pattern
composition and tensor products (parallel composition) of patterns.  We show
next that such pattern combinations reflect the corresponding combinations
of unitary operators.  An easy proof of universality, based on a family of
2-qubit patterns follows.

So far, this constitutes mostly a work of clarification of what was already
known from the series of papers introducing and investigating the
properties of the one-way model~\cite{mqqcs}.  However, we work here with
an extended notion of pattern, where inputs and outputs may overlap in any
way one wants them to, and this obtains more efficient - in the sense of
fewer qubits - implementations of unitaries.  Specifically, our generating
set consists of two simple patterns, each one using only 2 qubits.  From it
we obtain a 3 qubits realisation of the $R_z$ rotations and a 14 qubit
implementation for the controlled-$U$ family: a very significant reduction
over the known implementations.

However, the main point of this paper is to introduce alongside our
notation, a calculus of local equations over patterns that exploits the
fact that 1-qubit $xy$-measurements are closed under conjugation by Pauli
operators.  We show that this calculus is sound in that it preserves the
patterns interpretations.  Most importantly, we derive from it a simple
algorithm by which any general pattern can be put into a standard form
where entanglement is done first, then measurements, then corrections.

The consequences of the existence of such a procedure are far-reaching.
First, since entangling comes first, one can prepare the entire entangled
state needed during the computation right at the start: one never has to do
``on the fly'' entanglements.  Second, since local corrections come last,
only the output qubits will ever need corrections.  Third, the rewriting of
a pattern to standard form reveals parallelism in the pattern computation.
In a general pattern, one is forced to compute sequentially and obey
strictly the command sequence, whereas after standardisation, the
dependency structure is relaxed, resulting in low depth complexity.
Last, the existence of a standard form for any pattern also has interesting
corollaries beyond implementation and complexity matters, as it follows
from it that patterns using no dependencies, or using only the restricted class of Pauli measurements, can only realise a unitary
belonging to the Clifford group.

\mypar{Acknowledgements:}
Elham Kashefi wishes to express her gratitude to Quentin for letting her
collaborate with his father, Vincent Danos, during their stay in Canada.
Prakash Panangaden wishes to express his gratitude to EPSRC for supporting
his stay in Oxford where this collaboration began.

\section{Computation Patterns}
We first develop a notation for 1-qubit measurement based computations.  The
basic commands one can use are: 
\begin{itemize}
\item 1-qubit measurements $\M{\al}i$
\item 2-qubit entanglement operators $\et ij$
\item and 1-qubit Pauli corrections $\Cx i$, $\Cz i$
\end{itemize}
The indices $i$, $j$ represent the qubits on which each of these operations
apply, and $\al$ is a parameter in $[0,2\pi]$.  Sequences of such commands, 
together with two distinguished ---possibly overlapping--- sets of qubits
corresponding to inputs and outputs, will be called \emph{measurement
  patterns}, or simply patterns.  These patterns can be combined by
composition and tensor product.  

Importantly corrections and measurements are allowed to depend on previous
measurement outcomes.  We shall prove later that patterns without those
classical dependencies can only realise unitaries that are in the Clifford
group.  Thus dependencies are crucial if one wants to define a universal
computing model; that is to say a model where all finite-dimensional
unitaries can be realised, and it is also crucial to develop a notation
that will handle these dependencies gracefully.

\subsection{Commands}
The entanglement commands are defined as $\et ij:=\ctR Z_{ij}$,
while the correction commands are the Pauli operators $\Cx i$
and $\Cz i$.

A \emph{1-qubit measurement} command, written $\M{\al}i$, is given by a
pair of complement orthogonal projections, on: 
\EQ{
\oqb\al&:=&\ost(\ket0+ e^{i\al}\ket1)\\
\oqbn\al&:=&\ost(\ket0-\ei\al\ket1)
}
It is easily seen that $\oqb\al$, $\oqbn\al$ form
an orthonormal basis in $\ctwo$, so they indeed define a 1-qubit
measurement (of rank $2^{n-1}$, if $n$ is the number of qubits in the
ambient computing space).  
Measurements here will always be understood as destructive measurements,
that is to say the concerned qubit is consumed in the measurement
operation. 

The outcome of a measurement done at qubit $i$ will be denoted by
$s_i\in\ztwo$.  Since one only deals with patterns where qubits are
measured at most once (see condition (D1) below), this is unambiguous. We
take the convention that $s_i=0$ if under the corresponding measurement the
state collapses to $\oqb\al$, and $s_i=1$ if to $\oqbn\al$.

Outcomes can be summed together resulting in expressions of the form
$s=\sum_{i\in I} s_i$ which we call \emph{signals}, and where the summation
is understood as being done is $\ztwo$. We define the \emph{domain} of a
signal as the set of qubits it depends on. 

Dependent corrections will be written $\cx is$ and $\cz is$ with 
$s$ a signal. Their meaning is that $\cx i0=\cz i0=I$ (no
correction is applied), while $\cx i1=\Cx i$ and $\cz i1=\Cz i$.

Dependent measurements will be written $\MS{\al}ist$, where
$s$ and $t$ are signals. Their meaning is as follows:
\EQ{
\label{msem}
\MS{\al} ist&:=&\M{(-1)^s\al+t\pi} i
}
As a result, before applying a measurement, one has to know first all the measurements outcomes occurring in the signals $s$, $t$, then one has to compute the parity of $s$ and $t$, and maybe modify $\al$ to one of 
$-\al$, $\al+\pi$ and $-\al+\pi$.
One can easily compute that:
\EQ
{
X_i\M{{\al}}i X_i&=&\M{{-\al}}i\label{xmx}\\
Z_i\M{{\al}}i Z_i&=&\M{{\al+\pi}}i\label{zmz}
}
so that the actions correspond to conjugations of measurements under $X$
and $Z$. We will refer to them as the $X$ and $Z$-actions. Note that 
these two actions are commuting, since $-\al+\pi=-\al-\pi$ up to $2\pi$,
and hence the order in which one applies them doesn't matter.
Should one use other local corrections, then one would have here 
instead the corresponding actions on measurement angles. As we will see
later, relations (\ref{xmx}) and (\ref{zmz}) are key to the propagation of
dependent corrections, and to obtaining patterns in the standard
entanglement, measurement, correction form. Since measurements considered here are destructive ones, the equations simplify to $\M{{\al}}i X_i=\M{{-\al}}i$,
and $\M{{\al}}i Z_i=\M{{\al+\pi}}i$.

Another point worth noticing is that the domain of the signals of a
dependent command, be it a measurement or a correction, represents the set
of measurements which one has to do before one can determine the actual
value of the command.  

Finally we note that we could work with general 1-qubit measurements,
instead of the class defined above, sometimes called $xy$-measurements.
All the developments would carry through nicely, but we have not found so
far any compelling reason for this additional generality.

\subsection{Patterns}
\DE
Patterns consists of three finite sets $V$, $I$, $O$, together with two
injective maps $\iN:I\rar V$ and $\oU:O\rar V$ 
and a finite sequence of commands $A_n\ldots A_1$ applying to qubits in $V$. 
\ED
The set $V$ is called the pattern \emph{computation space}, and
we write $\hil V$ for the associated quantum state space $\otimes_{i\in V}\ctwo$. To ease notation, we will forget altogether about the
maps $\iN$ and $\oU$, and write simply $I$, $O$ instead of $\iN(I)$ and $\oU(O)$. Note however, that these maps are useful to define classical manipulations of the quantum states, such as permutations of the qubits. The sets 
$I$, $O$ will be called respectively the pattern \emph{inputs} and
\emph{outputs}, and we will write $\hil I$, and $\hil O$ for the associated
quantum state spaces. The sequence $A_n\ldots A_1$ will be called the pattern
\emph{command sequence}. 

To run a pattern, one prepares the input qubits in some input state
$\psi\in\hil I$,  while the non-input qubits are all set in the $\ket{+}$
state, then the commands are executed in sequence, and finally the result
of the pattern computation is some $\phi\in\hil O$. 
There might be qubits in the pattern, which are neither inputs nor outputs
qubits, and are used as auxiliary qubits during the computation. Usually
one tries to use as few of them as possible, since these participate to the
\emph{space complexity} of the computation. 

Note that one does not require inputs and outputs to be disjoint subsets of
$V$.  This seemingly innocuous additional flexibility is actually quite
useful to give parsimonious implementations of
unitaries~\cite{generator04}.  While the restriction to disjoint inputs and
outputs is unnecessary, it has been discussed whether more
constrained patterns might be easier to realise physically.  Recent
work~\cite{graphstates,BR04,CMJ04} however, seems to indicate they are not. 
\smallskip

Here is an example of a pattern implementing the Hadamard operator $H$:
\AR{
\mfr H&:=&
(\ens{1,2},\ens{1},\ens{2},\cx 2{s_1}\M01\et12)
}

What is this pattern doing~? The first qubit is prepared 
in some input state $\psi$, and the second in state $\ket+$, then these are
entangled to obtain $\ctR Z_{12}(\psi_1\otimes\ket+_2)$.  Once this is done,
the first qubit is measured in the $\ket+$, $\ket-$ basis.  Finally an $X$  
correction is applied on the output qubit, depending on the outcome of the
measurement.  We will do this calculation in detail later. 

\subsection{Pattern combination}
We are interested now in how one can combine patterns into bigger ones.

The first way to combine patterns is by composing them. 
Two patterns $\mfr P_{1}$ and $\mfr P_{2}$ may be composed if
$V_1\cap V_2=O_1=I_2$.
Note that provided that $\mfr P_1$ has as many outputs
as $\mfr P_2$ has inputs, by renaming the pattern qubits, one can always 
make them composable.
\DE
The composite pattern $\mfr P_{2}\mfr P_{1}$ is defined as:\\ 
--- $V:=V_1\cup V_2$, $I=I_1$, $O=O_2$,\\
--- commands are concatenated.
\ED
The other way of combining patterns is to tensor them.
Two patterns $\mfr P_{1}$ and $\mfr P_{2}$ 
may be tensored if $V_1\cap V_2=\emptyset$.
Again one can always meet this condition by renaming qubits
in a way that these sets are made disjoint. 
\DE
The tensor pattern $\mfr P_{2}\otimes\mfr P_{1}$ is defined as:\\
--- $V=V_1\cup V_2$, $I=I_1\cup I_2$, and $O=O_1\cup O_2$,\\ 
--- commands are concatenated. 
\ED
Note that all unions above are disjoint.
Note also that, in opposition to the composition case, commands from
distinct patterns freely commute, since they apply to disjoint qubits and
are independent of each other, so when we say that commands have to be
concatenated, this is only for definiteness. 

\subsection{Pattern conditions}
One might want to subject patterns to various conditions:
\begin{description}
\item[(D0)] no command depends an outcome not yet measured;
\item[(D1)] no command acts on a qubit already measured;
\item[(D2)] a qubit $i$ is measured if and only if $i$ is not an output;
\item[(EMC)] commands occur $E$s first, then $M$s, then $C$s. 
\end{description}
The reader might want to check that our example $\mfr H$ satisfies all of
the above.  It is routine to verify that these conditions are preserved
under composition and tensor.  Conditions (D0) and (D1) ensure that a
pattern can always be run meaningfully.  Indeed if (D0) fails, then at some
point of the computation, one will want to execute a command which 
depends on outcomes that are not known yet.  Likewise, if (D1) fails, one
will try to apply a command on a qubit that has been consumed by a
measurement (recall that we use destructive measurements). 
Condition (D2) is there to make sure that at the end of running the
pattern, the state will belong to the output space $\hil O$, \ie, that all
non-output qubits, and only them, will have been consumed by a measurement
when the computation ends. 

Starting now we will assume that all patterns satisfy the 
\emph{definiteness} conditions (D0), (D1) and (D2), and will designate by (D)
the conjunction of these three conditions. 

Condition (EMC) is of a completely different nature.
Patterns not respecting it will be called \emph{wild}. 

Later on, we will introduce the measurement calculus and show a simple
rewriting procedure turning any given wild pattern into an equivalent one
which is in (EMC) form.  We call this procedure \emph{standardisation}, and
also say that a pattern meeting the (EMC) condition is \emph{standard}. 

Before turning to this matter, we need a clean definition
of what it means for a pattern to implement or to realise a unitary operator,
together with a proof that the way one can combine patterns is reflected in
their interpretations.  This is key to our proof of universality. 

\section{Computing a pattern}
Besides quantum states which are vectors in some $\hil V$,
one needs a classical state recording the outcomes of the successive
measurements one does in a pattern. So it is natural to define the computation
state space as:
\AR{
\mcl S&:=&\bigcup_{V,W} \hil V\times\ztwo^W
}
where $V$, $W$ range over finite sets. In other words a computation state 
is a pair $q$, $\Ga$, where $q$ is a quantum state and $\Ga$ is a map from
some $W$ to the outcome space $\ztwo$. We call this classical component
$\Ga$ an \emph{outcome map} and denote by $\emptyset$ the unique map in
$\ztwo^\emptyset$. 

\subsection{Commands as actions}
We need a few notations. For any signal $s$ and classical state 
$\Ga\in\ztwo^W$, such that the domain of $s$ is included in $W$, we take
$s_\Ga$ 
to be the value of $s$ given by the outcome map $\Ga$. That is to say,
if $s=\sum_I s_i$, then $s_\Ga:=\sum_I\Ga(i)$ where the sum is taken in
$\ztwo$. 
Also if $\Ga\in\ztwo^W$, and $x\in\ztwo$, we define:
\AR{\Ga[x/i](i)=x,\,\Ga[x/i](j)=\Ga(j)\hbox{ for }j\neq i}
which is a map in $\ztwo^{W\cup\ens i}$.

We may now see each of our commands as acting on $\mcl S$.

\AR{
q,\Ga&\slar{\et ij}&\ctR Z_{ij} q,\Ga\\
q,\Ga&\slar{\cx is}&\cx i{s_\Ga} q,\Ga\\
q,\Ga&\slar{\cz is}&\cz i{s_\Ga} q,\Ga\\
q,\Ga&\slar{\MS{\al}ist}&{\oqbb{\al_\Ga}}_iq,\Ga[0/i]\\
q,\Ga&\slar{\MS{\al}ist}&{\oqbnb{\al_\Ga}}_iq,\Ga[1/i]
}
where $\al_\Ga=(-1)^{s_\Ga}\al+t_\Ga\pi$ following equation (\ref{msem}),
and $\bra\psi_i$ is the linear form associated to $\psi$ applied at
qubit $i$.
Suppose $q\in\hil V$, for the above relations to be defined, one needs the
indices $i$, $j$ on which the various command apply to be in $V$. One also
needs $\Ga$ to contain the domains of $s$ and $t$, so that $s_\Ga$ and
$t_\Ga$ are well-defined. This will always be the case during the run of a
pattern because of condition (D).

All commands except measurements are deterministic and only modify the
quantum part of the state. The measurements actions on $\mcl S$ are not
deterministic, so that these are actually binary relations on $\mcl S$, and
modify both the quantum and classical parts of the state. The usual
convention has it that when one does a measurement the resulting state is
\emph{renormalised}, but we don't adhere to it here, the reason being that
this way, the probability of reaching a given state can be read off its
norm, and the overall treatment is simpler.

We introduce an additional command called \emph{shifting}:
\AR{
q,\Ga&\slar{\ss is}&q,\Ga[\Ga(i)+s_\Ga/i]
}
It consists in shifting the
measurement outcome at $i$ by the amount $s_\Ga$. 
Note that the $Z$-action leaves measurements globally invariant, in the
sense that $\oqb{\al+\pi},\oqbn{\al+\pi}=\oqbn{\al},\oqb{\al}$.
Thus changing $\al$ to $\al+\pi$ amounts to swap the outcomes of the
measurements, and one has:
\EQ{\MS\al ist&=&\ss it\ms\al is\label{split}}
and shifting allows to split the $t$ action of a measurement,
resulting sometimes in convenient optimisations of standard forms.

\subsection{Computation branches}
Let $\mfr P$ be a pattern with computation space $V$, inputs
$I$, outputs $O$ and command sequence $A_n\ldots A_1$. A 
complete pattern computation starts with some input state $q$ in $\hil I$,
together with the empty outcome map $\emptyset$.  
The input state $q$ is then tensored with as many $\ket+$s as there are
non-inputs in $V$, so as to obtain a state in the full space $\hil V$. 
Then commands in $\mfr P$ are applied in sequence. We can summarise the
situation as follows: 
\AR{
\xymatrix@=10pt@M=5pt@R=20pt@C=40pt{
{}\hil I\ar[d]\ar@{.>}[rr]
&&
{}\hil O
\\
{}\hil I\times\ztwo^{\emptyset}\ar[r]^{prep}&
{}\hil V\times\ztwo^{\emptyset}\ar[r]^{A_1\ldots A_n\quad}&
{}\hil O\times\ztwo^{V\setminus O}\ar[u]
}
}
To make this precise, say there is a \emph{$\mfr P$-branch} from
$q\in\hil I$ to $q'\in\hil O$, written $q \brA{\mfr P}q'$, 
if there is a sequence 
$(q_i,\Ga_i)$ with $1\leq i\leq n+1$, such that: 
\AR{
q\otimes\ket{\hskip-.4ex+\ldots+},\emptyset=q_1,\Ga_1\\ 
q'=q_{n+1}\neq0\\
\hbox{and for all }i\leq n:q_i,\Ga_i\slar{A_i}q_{i+1},\Ga_{i+1}
}
thus $\brA{\mfr P}$ is a binary relation on 
$\hil I\times\hil O$. That it is a relation and not a map
reflects the fact that measurements a priori introduce non determinism
in the evolution of the quantum states.

Specifically, if $k$ is the number of measurements in $\mfr P$ (or
equivalently the number of non-outputs qubits), there are at most $2^k$
branches in any given computation, and therefore a given $q\in\hil I$ is in
relation with at most 
$2^k$ distinct $q'\in\hil O$. The \emph{probability} of a branch is  
defined to be $\|q'\|^2/\|q\|^2$ ($q$ being always assumed to be non zero).
Indeed one has:
\EQ{
\label{likeli}
\sum_{\ens{q'\mid q \brA{\mfr P}q'}}\|q'\|^2&=&\|q\|^2
}
since any action is either a unitary, thus a norm-preserving action,
or a measurement which introduces a branching, and then if 
$q$ projects to $q_0$ and $q_1$, under some 
$\M\al i$, $\|q\|^2=\|q_0\|^2+\|q_1\|^2$, 
so that the relation above is always preserved.

\DE
One says the pattern $\mfr P$ is \emph{{deterministic}}
if for all $q\in\hil I$, $q'$ and $q''\in\hil O$, whenever
$q\brA{\mfr P}q'$ and $q\brA{\mfr P}q''$,
then $q'$ and $q''$ only differ up to a scalar.
\ED
Note that even when $\mfr P$ is deterministic, all branches
might not be equally likely.
When $\mfr P$ is deterministic, 
one defines a norm-preserving map $U_{\mfr  P}$ from 
$\hil I$ to $\hil O$ by:
\EQ{
U_{\mfr  P}(q)&:=&\frac{\|q\|}{\|q'\|}q'
}
Note that when $q\brA{\mfr P}q'$, $q'\neq0$, so that the definition above
always make sense. Note also that because $\mfr P$ is deterministic, this
map depends on the choice of $q'$ only up to a global phase. One can
further comment that 
since we took the convention not to renormalise measurement results,
we have to do here a global renormalisation to define the pattern
interpretation.

One says that a deterministic pattern $\mfr P$ \emph{realises} or
\emph{implements} $U_{\mfr  P}$, or equivalently that $U_{\mfr  P}$ is the
\emph{interpretation} of $\mfr P$.  

This map $U_{\mfr P}$ must actually be a unitary embedding, since all
quantum definable deterministic transformations are unitaries.  If a
precise argument is needed here, one can rephrase all the definitions given
so far in the language of density operators and completely-positive maps
(cp-maps).  Then a deterministic pattern will implement a cp-map preserving
pure density operators.  From the Kraus representation theorem for cp-maps,
it is easy to see that such cp-maps are liftings of unitary embeddings.

\subsection{Short examples}
First we give a quick example of a deterministic pattern that has branches
with different probabilities. The state space is $\ens{1,2}$,
with $I=O=\ens1$, while the command sequence is $\M\al2$.
Therefore, starting with input $q$, one gets two branches:
\AR{
q\otimes\ket+,\emptyset
&\slar{\M\al2}&
\left\{
\begin{array}{l}
\frac12(1+\emi{\al})q,\emptyset[0/2]\\\\
\frac12(1-\emi{\al})q,\emptyset[1/2]
\end{array}
\right.
}
Thus this pattern is indeed deterministic, and
implements the identity up to a global phase, 
and yet the two branches have respective probabilities
$(1+\cos\al)/2$ and $(1-\cos\al)/2$, which are not equal in 
general.

Next, we return to the pattern $\mfr H$ which we already took as an
example. Let us consider for a start the pattern with same space
$\ens{1,2}$, same inputs and outputs $I=\ens1$, $O=\ens2$, and shorter
command sequence $\Ms 01\et12$. 
Starting with input $q=(a\ket{0}+b\ket{1})\ket+$, one has two computation
branches, branching at $\Ms 01$:  
\AR{
(a\ket{0}+b\ket{1})\ket+,\emptyset
&\slar{\et12}&
\ost(a\ket{00}+a\ket{01}+b\ket{10}-b\ket{11}),\emptyset\\\\
&\slar{\Ms 01}& 
\left\{
\begin{array}{l}
\frac12((a+b)\ket{0}+(a-b)\ket{1}),\emptyset[0/0]\\\\
\frac12((a-b)\ket{0}+(a+b)\ket{1}),\emptyset[1/0]
\end{array}
\right.
}
and since $\norm{a+b}^2+\norm{a-b}^2=2(\norm a^2+\norm b^2)$,
both transitions happen with equal probabilities $\frac12$. 
Both branches end up with different outputs, so the pattern is \emph{not}
deterministic. However, if one applies the local correction $\Cx2$ on
either of the branches ends, both outputs will be made to coincide. Let us
choose to let the correction bear on the second branch, obtaining the
example $\mfr H$ 
which we defined already. We have just proved $H=U_{\mfr H}$,
that is to say $\mfr H$ realises the Hadamard operator.

With our definitions in place, we first infer that 
pattern combinations correspond to combinations of
their interpretations.  From this an easy structured argument - that uses
surprisingly simple patterns - for universality will follow. 

\subsection{Composing, Tensoring and Interpretation}
Recall that two patterns $\mfr P_1$, $\mfr P_2$ may be combined by
composition provided $\mfr P_1$ have as many outputs as $\mfr P_2$ has
inputs. Suppose 
this is the case, and suppose further that $\mfr P_1$ and $\mfr P_2$
respectively realise some unitaries $U_1$ and $U_2$, then  
the composite pattern $\mfr P_2\mfr P_1$ realises $U_2U_1$.

Indeed, the two diagrams representing branches in $\mfr P_1$
and $\mfr P_2$:

{\footnotesize
\AR{
\xymatrix@=10pt@M=3pt@R=20pt@C=7pt{
{}\hil {I_1}\ar[d]\ar@{.>}[rr]
&&
{}\hil {O_1}\ar@{=}
&
{}\hil {I_2}\ar[d]\ar@{.>}[rr]
&&
{}\hil {O_2}
\\
{}\hil {I_1}\times\ztwo^{\emptyset}\ar[r]^{p_1}&
{}\hil {V_1}\times\ztwo^{\emptyset}\ar[r]^{}&
{}\hil {O_1}\times\ztwo^{V_1\setminus O_1}\ar[u]
&
{}\hil {I_2}\times\ztwo^{\emptyset}\ar[r]^{p_2}&
{}\hil {V_2}\times\ztwo^{\emptyset}\ar[r]^{}&
{}\hil {O_2}\times\ztwo^{V_2\setminus O_2}\ar[u]
}
}
}\\
can be pasted together, since $O_1=I_2$, and $\hil {O_1}=\hil {I_2}$. 
But then, it is enough to notice 1) that preparation steps $p_2$ in $\mfr P_2$
commute to all actions in $\mfr P_1$ since they apply on disjoint sets
of qubits, and 2) that no action taken in $\mfr P_2$ depends on 
the measurements outcomes in $\mfr P_1$. It follows that 
the pasted diagram describes the same branches as does
the one associated to the composite $\mfr P_2\mfr P_1$. 

A similar argument applies to the case of a tensor combination,
and one has that $\mfr P_2\otimes\mfr P_1$ realises $U_2\otimes U_1$.

The same holds even for non-deterministic patterns considered as
implementing cp-maps.  But we will not be concerned with this generalised
setting in this paper.

\section{Universality}
Consider the two following patterns: 
\EQ{
\mfr \G(\al)&:=&\cx 2{s_1}\M{{-\al}}1\et 12\\
\ctR{\mfr Z}&:=&\et 12 
}
In the first pattern $1$ is the only input and $2$ is the only output,
while in the second both $1$ and $2$ are inputs and outputs. Note that here
we are taking advantage of allowing patterns with overlapping inputs and
outputs. 

\PRO
The patterns $\mfr \G(\al)$ and $\ctR{\mfr Z}$ are universal.
\ORP
First, we claim $\mfr \G(\al)$ and $\ctR{\mfr Z}$ respectively
realise $\G(\al)$ and $\ctR{Z}$, with:
\AR{ 
\G(\al)&:=&\ost\MA{1&\ei\al\\1&-\ei\al}
}
We have already seen in our example that $\mfr\G(0)=\mfr H$ implements
$H=\G(0)$, thus we already know this in the particular 
case where $\al=0$. The general case follows by the same kind of computation.
The case of $\ctR Z$ is obvious.\\
Second, we know that these unitaries form a universal
set~\cite{generator04}. Therefore, from the preceding section, 
we infer that combining the corresponding patterns
will generate patterns realising all finite-dimensional unitaries.
\qed

These patterns are indeed among the simplest possible. As a consequence, in the section devoted to examples, we will find that our implementations have often little space complexity.

Remarkably, in our set of generators, one finds a single dependency, which occurs in the correction phase of $\mfr \G(\al)$.  No set of patterns without any measurement could be a generating set, since such patterns can only implement unitaries in the Clifford group.  Dependencies are also needed for universality, but we have to wait for the development of the measurement calculus in the next section to give a proof of this fact.

\section{The measurement calculus}
We turn to the next important matter of the paper, namely standardisation.
The idea is quite simple. It is enough to provide local pattern rewriting
rules pushing $E$s to the beginning of the pattern, and $C$s to the end.  

\subsection{The equations}
A first set of equations give means to propagate local Pauli corrections
through the entangling operator $\et ij$. Because $\et ij =\et ji$, there are only two cases to consider: 
\EQ{
\et ij \cx is&=&\cx is\cz js\et ij\label{ecx}\\
\et ij \cz is&=&\cz is\et ij\label{ecz}
}
These equations are easy to verify and are natural since
$\et ij$ belongs to the Clifford group, and therefore maps 
under conjugation the Pauli group to itself. 

A second set of equations give means to push corrections through
measurements acting on the same qubit. Again there are two cases: 
\EQ{ 
\MS\al ist\cx ir&=&\MS\al i{s+r}{t}\label{mcx}\\
\MS\al ist\cz ir&=&\MS\al i{s}{t+r}\label{mcz}
}
These equations follow easily from equations (\ref{xmx}) and (\ref{zmz}).
They express the fact that the 
measurements $\Ms\al i$ are closed under conjugation by the Pauli group, 
very much like equations~(\ref{ecx}) 
and~(\ref{ecz}) express the fact that the Pauli group is closed under
conjugation by the entanglements $\et ij$.

Define the following convenient abbreviations:
\AR{
\ms\al is:=\MS\al is0,\,
\MS\al i{}t:=\MS\al i0t,\,
\Ms\al i:=\MS\al i00,\\
\Ms xi:=\Ms0i,\,
\Ms yi:=\Ms\pit i
}
Particular cases of the equations above are:
\AR{ 
\M xi\cx is&=&\M xi\\
\M yi\cx is&=&\ms yis &=&\MS yi{}s &=&\M yi\cz is
}
The first equation, follows from ${-0}=0$, so the $X$ action on $\M xi$
is trivial; the middle equation, second row, is because 
${-\pit}$ is equal $\pit+\pi$ modulo $2\pi$, 
and therefore the $X$ and $Z$ actions coincide on $\M yi$.
So we obtain the following:
\EQ{ 
\MS xist&=&\MS xi{}t\label{mx}\\
\MS yist&=&\MS yi{}{s+t}\label{my}
}
which we will use later to prove that patterns with measurements
of the form $M^x$ and $M^y$ may only realise unitaries in the 
Clifford group.

\subsection{The rewriting rules}
We now define a set of rewrite rules, obtained by directing the equations
above:  
\AR{
\et ij\cx is&\Rar&\cx is\cz js\et ij&\quad\hbox{}EX\\
\et ij\cz is&\Rar&\cz is\et ij&\quad\hbox{}EZ\\
\MS\al is{t}\cx i{r}&\Rar&\MS\al i{s+r}{t}&\quad\hbox{}MX\\
\MS\al is{t}\cz i{r}&\Rar&\MS\al i{s}{r+t}&\quad\hbox{}MZ }
to which we need to add the \emph{free commutation rules}, 
obtained when commands operate on disjoint sets of qubits: 
\AR{
\et ij\CO{\vec k}&\Rar&\CO{\vec k}\et ij&\quad\hbox{with }A\neq E\\
\CO{\vec k}\cx is&\Rar&\cx is\CO{\vec k}&\quad\hbox{with }A\neq C\\
\CO{\vec k}\cz is&\Rar&\cz is\CO{\vec k}&\quad\hbox{with }A\neq C 
} 
where $\vec k$ represent the qubits acted upon by command $A$,
and are supposed to be distinct from $i$ and $j$.

Condition (D) is easily seen to be preserved under rewriting.

Under rewriting, the computation space, inputs and outputs remain the same, and so are the entanglement commands. Measurements might be modified, but there is still the same number of them, and they are still acting on the
same qubits. The only induced modifications concern local corrections and
dependencies. We also take due note that none of these equations may create 
dependencies.

\subsection{Standardisation}
Write $\mfr P\Rar\mfr P'$, respectively $\mfr P\Rar\st\mfr P'$, if both
patterns have the same type, and one obtains $\mfr P'$'s command 
sequence from $\mfr P$'s one by applying one, respectively any number, of
the rules above. Say $\mfr P$ is \emph{standard} if for no $\mfr P'$, $\mfr
P\Rar\mfr P'$. 

Because all our equations are sound, one has that whenever $\mfr
P\Rar\st\mfr P'$, and both patterns are deterministic, then $U_{\mfr
  P}=U_{\mfr P'}$.

One can show by a standard rewriting theory argument, 
that for all $\mfr P$, there exists a unique standard $\mfr P'$, such that 
$\mfr P\Rar\st\mfr P'$, and moreover $\mfr P'$ satisfies the (EMC) condition. 
Reaching the standard form takes at most quadratic time
in the number of instructions in $\mfr P$.  Details are given in the appendix. 

\subsection{Signal shifting}
One can extend the calculus to include the shifting command $\ss it$.
This allows one to dispose of dependencies induced by the $Z$-action, and
obtain sometimes standard patterns with smaller depth complexity, as we will see in the next section devoted to examples. 
\EQ{
\MS\al ist&\Rar&\ss it\ms\al is\\
\cx js\ss it&\Rar& \ss it \cx j{s[t+s_i/s_i]}\\
\cz js\ss it&\Rar& \ss it \cz j{s[t+s_i/s_i]}\\
\MS\al jst\ss ir&\Rar&\ss ir \MS\al j{s[r+s_i/s_i]}{t[r+s_i/s_i]}
}
where $s[t/s_i]$ is the substitution of $s_i$ with
$t$ in $s$, $s$, $t$ being signals.
The first additional rewrite rule was already introduced
as equation (\ref{split}), while the other ones are merely propagating
the signal shift. Clearly also, one can dispose of $\ss it$ when it hits
the end of the pattern command sequence. We will refer to this new set of
rules as $\Rar_S$.

\section{Examples}
In this section we develop some examples illustrating both pattern
composition, pattern standardisation, and signal shifting. We compare our
implementations with the implementations given in the reference
paper~\cite{mqqcs}. To combine patterns 
one needs to rename their qubits as we already noticed. We use the 
following concrete notation: if $\mfr P$ is a pattern over
$\ens{1,\ldots,n}$,  
and $f$ is an injection, we write $\mfr P(f(1),\ldots,f(n))$
for the same pattern with qubits renamed according to $f$. We also
write $\mfr P_2\circ\mfr P_1$ for pattern composition to ease reading.

\subsubsection*{Teleportation.}
Consider the composite pattern $\mfr\G(\ba)(2,3)\circ\mfr\G(\al)(1,2)$ with
computation space $\ens{1,2,3}$, inputs $\ens{1}$, and outputs $\ens{3}$. 
We run our standardisation procedure so as to obtain an equivalent standard
pattern:  
\AR{
\mfr \G(\ba)(2,3)\circ\mfr \G(\al)(1,2)&=&
\cx3{s_{2}}\Ms{-\ba}2\tr{\et23\cx2{s_{1}}}\Ms{-\al}1\et12
\\&\Rar_{EX}&
\cx3{s_{2}}\tr{\Ms{-\ba}2\cx2{s_1}}\cz3{s_1}\Ms{-\al}1\et23\et12
\\&\Rar_{MX}&
\cx3{s_{2}}\cz3{s_{1}}\ms{-\ba}2{s_{1}}\Ms{-\al}1\et23\et12
}
Let us call the pattern just obtained $\mfr \G(\al,\ba)$. If we take as
a special case $\al=\ba=0$, we get: 
\AR{
\cx3{s_2}\cz3{s_1}\Ms x2\Ms x1\et23\et12
}
and since we know that $\mfr \G(0)$ implements $H$ and $H^2=I$, we conclude
that this pattern 
implements the identity, or in other words it teleports qubit $1$ to
qubit $3$.  As it happens, this pattern obtained by self-composition, is the
same as the one given in the reference paper~\cite[p.14]{mqqcs}. 

\subsubsection*{$x$-rotation.}
Here is the reference implementation of an $x$-rotation~\cite[p.17]{mqqcs},
$R_x(\al)$: 
\EQ{
\cx3{s_2}\cz3{s_1}\ms{-\al}2{s_1}\Ms x1\et23\et12
}
with computation space $V=\ens{1,2,3},\ens{1},\ens{3}$.
There is a natural question which me might call the 
recognition problem, namely how do we know this is implementing
$R_x(\al)$~? Of course there is the brute force answer to that, which we
applied to compute our simpler patterns, and which consists in computing
down all the four possible branches  
generated by the measurements at $1$ and $2$.  Another possibility is to use
the stabiliser formalism as explained in the reference paper~\cite{mqqcs}. 
Yet another possibility is to use \emph{pattern composition}, as we did
before, and this is what we are going to do. 

We know that $R_x(\al)=\G({\al})H$ up to a global phase, hence the
composite pattern $\mfr \G({\al})(2,3)\circ\mfr H(1,2)$ implements
$R_{x}(\al)$. 
Now we may standardise it:
\AR{
\mfr \G({\al})(2,3)\circ\mfr H(1,2)&=&
\cx3{s_2}\Ms{-\al}2\tr{\et23\cx2{s_1}}\Ms x1\et12\\
&\Rar_{EX}&
\cx3{s_2}\cz3{s_1}\tr{\Ms{-\al}2\cx2{s_1}}\Ms x1\et23\et12\\
&\Rar_{MX}&
\cx3{s_2}\cz3{s_1}\ms{-\al}2{s_1}\Ms x1\et23\et12\\
}
obtaining exactly the implementation we started with.
Since our calculus is preserving interpretations, we deduce that
the implementation is correct.

\subsubsection*{$z$-rotation.}
Now, we have a method here for synthesising further implementations,
which we can use fir instance with another rotation $R_z(\al)$. 
Again we know that $R_z(\al)=HR_x(\al)H$, and we already know how to
implement both components $H$ and $R_x(\al)$.  

Starting with the pattern
$\mfr H(4,5)\circ\mfr R_x(\al)(2,3,4)\circ\mfr H(1,2)$ we get:
\AR{
\mfr H(4,5)\circ\mfr R_x(\al)(2,3,4)\circ\mfr H(1,2)=\\
\mfr H(4,5)
\cx4{s_3}\cz4{s_2}
\MS\al3{1+s_2}{}
\Ms x2
\et34
\tr{\et23
\cx2{s_1}}
\Ms x1
\et12
&\Rar_{EX}&\\
\mfr H(4,5)
\cx4{s_3}\cz4{s_2}\MS\al3{1+s_2}{}\Ms x2
\cx2{s_1}
\tr{\et34\cz3{s_1}}
\Ms x1
\et1{23}
&\Rar_{EZ}&\\
\mfr H(4,5)
\cx4{s_3}\cz4{s_2}\MS\al3{1+s_2}{}\cz3{s_1}
\tr{\Ms x2\cx2{s_1}}
\Ms x1
\et1{234}
&\Rar_{MX}&\\
\mfr H(4,5)
\cx4{s_3}\cz4{s_2}
\tr{\MS\al3{1+s_2}{}\cz3{s_1}}
\Ms x2
\Ms x1
\et1{234}
&\Rar_{MZ}&\\
\cx5{s_4}
\Ms x4
\tr{\et45\cx4{s_3}}\cz4{s_2}
\MS\al3{1+s_2}{s_1}
\Ms x2
\Ms x1
\et1{234}
&\Rar_{EX}&\\
\cx5{s_4}
\cz5{s_3}
\tr{\Ms x4\cx4{s_3}}
\cz4{s_2}
\MS\al3{1+s_2}{s_1}
\Ms x2
\Ms x1
\et1{2345}
&\Rar_{MX}&\\
\cx5{s_4}
\cz5{s_3}
\tr{\MS x4{s_3}{}\cz4{s_2}}
\MS\al3{1+s_2}{s_1}
\Ms x2
\Ms x1
\et1{2345}
&\Rar_{MZ}&\\
\cx5{s_4}
\cz5{s_3}
\MS x4{s_3}{s_2}
\MS\al3{1+s_2}{s_1}
\Ms x2
\Ms x1
\et1{2345}
}
To ease reading $\et23\et12$ is shortened to $\et1{23}$, $\et12\et23\et34$
to $\et1{234}$, and $\MS{\al}i{1+s}{t}$ is used as shorthand for
$\MS{-\al}i{s}{t}$.  

Here for the first time, we see $MZ$ rewritings, inducing the $Z$-action on
measurements. The obtained standardised pattern can therefore be rewritten
further using the extended calculus: 
\AR{
\cx5{s_4}
\cz5{s_3}
\MS x4{s_3}{s_2}
\MS\al3{1+s_2}{s_1}
\Ms x2
\Ms x1
\et1{2345}
&\Rar_{S}&\\
\cx5{s_2+s_4}\cz5{s_1+s_3}
\Ms x4
\MS\al3{1+s_2}{}
\Ms x2\Ms x1
\et1{2345}
}
obtaining again the pattern given in the reference paper~\cite[p.5]{mqqcs}.

However, just as in the case of the $R_x$ rotation, we also
have $R_z(\al)=H\G({\al})$ up to a global phase, 
hence the pattern $\mfr H(2,3)\mfr \G({\al})(1,2)$ also 
implements $R_{z}(\al)$, and we may standardize it:
\AR{
\mfr H(2,3)\circ\mfr \G({\al})(1,2)&=&
\cx3{s_2}\Ms x2
\tr{\et23\cx2{s_1}}
\Ms{-\al}1\et12
\\&\Rar_{EX}&
\cx3{s_2}
\cz3{s_1}
\tr{\Ms x2
\cx2{s_1}
}\Ms{-\al}1\et1{23}
\\&\Rar_{MX}&
\cx3{s_2}
\cz3{s_1}
\Ms x2{}{}
\Ms{-\al}1\et1{23}
}
obtaining a 3 qubits standard pattern for the $z$-rotation, which is simpler
than the preceding one, because it is based on the $\mfr \G(\al)$ generators. 
Since the $z$-rotation $R_z(\al)$ is the same as the phase operator:
\AR{P(\al)=\MA{1&0\\0&\ei\al}}
up to a global phase, we also obtain with the same pattern an implementation
of the phase operator. In particular, if $\al=\pit$, using
the extended calculus, we get the following pattern for $P(\pit)$:
$\cx3{s_2}\cz3{s_1+1}\Ms x2\Ms y1\et1{23}$.

\subsubsection*{General rotation.}
The realisation of a general rotation based
on the Euler decomposition of rotations as $R_x(\ga)R_z(\ba)R_x(\al)$,
would results in a 7 qubits pattern. We get a 5 qubits implementation 
based on the $\G(\al)$ decomposition~\cite{generator04}:
\AR{
R(\al,\ba,\ga)&=&\G(0) \G(\al) \G(\ba) \G(\ga)
}
The extended standardization procedure yields:
\AR{
\mfr \G(0)(4,5)
\mfr \G(\al)(3,4)
\mfr \G(\ba)(2,3)
\mfr \G(\ga)(1,2)
&=&\\
\cx5{s_4}\Ms{0}4\et45
\cx4{s_3}\Ms{\al}3\et34
\cx3{s_2}\Ms{\ba}2
\tr {\et23 \cx2{s_1}}
\Ms{\ga}1\et12
&\Rar_{EX}&\\
\cx5{s_4}\Ms{0}4\et45
\cx4{s_3}\Ms{\al}3\et34
\cx3{s_2}
\tr{\Ms{\ba}2\cx2{s_1}}
\cz3{s_1}
\Ms{\ga}1\et1{23}
&\Rar_{MX}&\\
\cx5{s_4}\Ms{0}4\et45
\cx4{s_3}\Ms{\al}3
\tr{\et34\cx3{s_2}\cz3{s_1}}
\ms{\ba}2{s_1}
\Ms{\ga}1\et1{23}
&\Rar_{EXZ}&\\
\cx5{s_4}\Ms{0}4\et45
\cx4{s_3}
\tr{\Ms{\al}3\cx3{s_2}\cz3{s_1}}
\cz4{s_2}
\ms{\ba}2{s_1}
\Ms{\ga}1\et1{234}
&\Rar_{MXZ}&\\
\cx5{s_4}\Ms{0}4
\tr{\et45\cx4{s_3}\cz4{s_2}}
\MS{\al}3{s_2}{s_1}
\ms{\ba}2{s_1}
\Ms{\ga}1\et1{234}
&\Rar_{EXZ}&\\
\cx5{s_4}
\tr{\Ms{0}4\cx4{s_3}\cz4{s_2}}
\cz5{s_3}
\MS{\al}3{s_2}{s_1}
\ms{\ba}2{s_1}
\Ms{\ga}1\et1{2345}
&\Rar_{MXZ}&\\
\cx5{s_4}\cz5{s_3}
\MS{0}4{}{s_2}
\MS{\al}3{s_2}{s_1}
\ms{\ba}2{s_1}
\Ms{\ga}1\et1{2345}
&\Rar_{S}&\\
\cx5{s_2+s_4}\cz5{s_1+s_3}
\Ms{0}4
\MS{\al}3{s_2}{}
\ms{\ba}2{s_1}
\Ms{\ga}1\et1{2345}
}

\subsubsection*{CNOT ($\ctR X$).}
This is our first example with two inputs and two outputs.
We use here the trivial pattern $\mfr I$ with computation space $\ens1$, 
inputs $\ens1$, outputs $\ens1$, and empty command sequence,
which implements the identity over $\hil 1$. 

One has $\ctR X=(I\otimes H)\ctR Z(I\otimes H)$, so we get a pattern using 
4 qubits over $\ens{1,2,3,4}$, with inputs $\ens{1,2}$, and outputs
$\ens{1,4}$, 
where one notices that inputs and outputs intersect on the control qubit 
$\ens1$:
\AR{
(\mfr I(1)\otimes\mfr h(3,4))
\ctR\mfr Z(1,3)
(\mfr I(1)\otimes\mfr h(2,3))
&=&
\cx4{s_3}
\Ms x3
\et34
\et13
\cx3{s_2}
\Ms x2
\et23
}
By standardising:
\AR{
\cx4{s_3}
\Ms x3
\et34
\tr{\et13\cx3{s_2}}
\Ms x2
\et23
&\Rar_{EX}&\\
\cx4{s_3}
\cz1{s_2}
\Ms x3
\tr{\et34\cx3{s_2}}
\Ms x2
\et13\et23
&\Rar_{EX}&\\
\cx4{s_3}
\cz4{s_2}
\cz1{s_2}
\tr{\Ms x3\cx3{s_2}}
\Ms x2
\et13\et23\et34
&\Rar_{MX}&\\
\cx4{s_3}
\cz4{s_2}
\cz1{s_2}
\Ms x3
\Ms x2
\et13\et23\et34
}
Note that we are not using here the $\et1{234}$ abbreviation, because
the underlying structure of entanglement is not a chain.
This pattern was already described in Aliferis and Leung's paper~\cite{AL04}.
In their original presentation the authors actually use an explicit
identity pattern (using the teleportation pattern $\mfr \G(0,0)$ presented
above), but we know from the careful presentation of composition
that this is not necessary.

\subsubsection*{GHZ.}
We present now a family of patterns preparing the GHZ entangled states
$\ket{0\ldots0}+\ket{1\ldots1}$. One has:
\AR{
\hbox{GHZ}(n)&=&
(H_n
\ctR Z_{n-1 n}
\ldots
H_2
\ctR Z_{1 2})\ket{\hskip-.4ex+\hskip-.4ex\ldots\hskip-.4ex+}
}
and by combining the patterns for $\ctR Z$ and $H$, we obtain
a pattern with computation space $\ens{1,2,2',\ldots, n, n'}$,
no inputs, outputs $\ens{1,2',\ldots,n'}$,
and the following command sequence:
\AR{
\cx{n'}{s_n}\Ms x{n}\et{n}{n'}
\et{(n-1)'}{n}
\ldots
\cx{2'}{s_2}\Ms x{2}\et{2}{2'}
\et{1}{2}
}
Under that form, the only apparent way to run the pattern is to execute
all commands in sequence. The situation changes completely, when we bring
the pattern to extended standard form:
\AR{
\cx{n'}{s_n}
\Ms x{n}\et{n}{n'}
\et{(n-1)'}{n}
\ldots
\cx{3'}{s_3}
\Ms x{3}
\et{3}{3'}
\tr{\et{2'}{3}
\cx{2'}{s_2}
}\Ms x{2}\et{2}{2'}
\et{1}{2}
&\Rar&\\
\cx{n'}{s_n}
\cx{2'}{s_2}
\Ms x{n}\et{n}{n'}
\et{(n-1)'}{n}
\ldots
\cx{3'}{s_3}
\tr{\Ms x{3}\cz{3}{s_2}}
\Ms x{2}
\et{3}{3'}
\et{2'}{3}
\et{2}{2'}
\et{1}{2}
&\Rar&\\
\cx{n'}{s_n}
\cx{2'}{s_2}
\Ms x{n}\et{n}{n'}
\et{(n-1)'}{n}
\ldots
\cx{3'}{s_3}
\MS x{3}{}{s_2}
\Ms x{2}
\et{3}{3'}
\et{2'}{3}
\et{2}{2'}
\et{1}{2}
&\Rar\st&\\
\cx{n'}{s_n}
\ldots
\cx{3'}{s_3}
\cx{2'}{s_2}
\MS x{n}{}{s_{n-1}}
\ldots
\MS x{3}{}{s_2}
\Ms x{2}
\et{n}{n'}
\et{(n-1)'}{n}
\ldots
\et{3}{3'}
\et{2'}{3}
\et{2}{2'}
\et{1}{2}
&\Rar_S&\\
\cx{n'}{s_2+s_3+\cdots+s_n}
\ldots
\cx{3'}{s_2+s_3}
\cx{2'}{s_2}
\Ms x{n}
\ldots
\Ms x{3}
\Ms x{2}
\et{n}{n'}
\et{(n-1)'}{n}
\ldots
\et{3}{3'}
\et{2'}{3}
\et{2}{2'}
\et{1}{2}
}
All measurements are now independent of each other, it is therefore
possible after the entanglement phase, to do all of them in one round, and
in a subsequent round to do all local corrections. In other words, the
obtained pattern has constant depth complexity $2$. 

\subsubsection*{Controlled-$U$}
\begin{figure}[h]
\begin{center}
\includegraphics[scale=0.6]{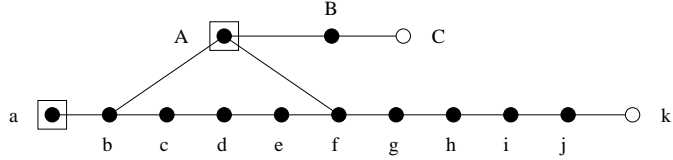}
\caption{The underlying entanglement graph for the $\ctR U$ pattern.}
\label{wildgraph}
\end{center}
\end{figure}

\begin{figure}[h]
\begin{center}
\includegraphics[scale=0.6]{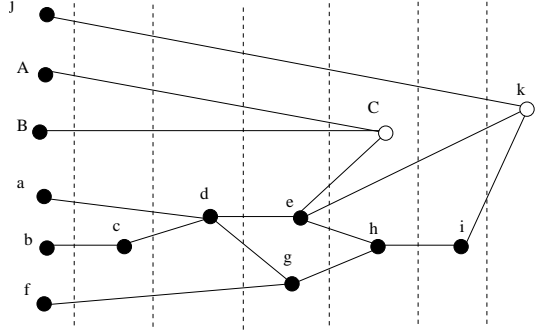}
\caption{The dependency graph for the standard $\ctR U$ pattern.}
\label{dependgraph}
\end{center}
\end{figure}

This final example presents another instance where standardization obtains
a low depth complexity.  
For any 1-qubit unitary $U$, one has the following decomposition
of $\ctR U$ in terms of the generators $\G(\al)$~\cite{generator04}:
\AR{ 
\ctR U_{12}
&=& 
\Rr{1}{0}\Rr{1}{\al'}
\Rr{2}{0}\Rr{2}{\ba+\pi}
\Rr2{-\frac\ga2}\Rr2{-\pit}
\Rr20\ctR Z_{12}
\Rr2{\pit}\Rr2{\frac\ga2}
\Rr2{\frac{-\pi-\da-\ba}2} 
\Rr20
\ctR Z_{12}
\Rr2{\frac{-\ba+\da-\pi}2} 
}
with $\al'=\al+\frac{\ba+\ga+\da}2$.
By translating each $\G$ operator to its corresponding pattern, we get the
following wild pattern for $\ctR U$: 
\AR{
\cx{C}{s_B}\Ms{0}{B}\et{B}{C}
\cx{B}{s_A}\Ms{-\al'}{A}\et{A}{B}
\cx{k}{s_j}\Ms{0}{j}\et{j}{k}
\cx{j}{s_i}\Ms{-\ba-\pi}{i}\et{i}{j}
\\
\cx{i}{s_h}\Ms{\frac\ga2}{h}\et{h}{i}
\cx{h}{s_g}\Ms{\pit}{g}\et{g}{h}
\cx{g}{s_f}\Ms{0}{f}\et{f}{g}
\et{A}f
\cx{f}{s_e}\Ms{-\pit}{e}\et{e}{f}
\\
\cx{e}{s_d}\Ms{-\frac\ga2}{d}\et{d}{e}
\cx{d}{s_c}\Ms{\frac{\pi+\da+\ba}2}{c}\et{c}{d}
\cx{c}{s_b}\Ms{0}{b}\et{b}{c}
\et{A}b
\cx{b}{s_a}\Ms{\frac{\ba-\da+\pi}2}{a}\et{a}{b}
}
Figure~\ref{wildgraph} shows the underlying entanglement graph for the
$\ctR U$ pattern. In order to run the wild form of the pattern one needs to
follow the graph structure and hence one has to perform the measurement
commands in sequence. 
Extended standardisation yields:
\AR{
\cz{k}{s_i+s_g+s_e+s_c+s_a}
\cx{k}{s_j+s_h+s_f+s_d+s_b}
\cx{C}{s_B}
\cz{C}{s_A+s_e+s_c}
\\
\Ms{0}{B}
\Ms{-\al'}{A}
\Ms{0}{j}
\ms{\ba-\pi}{i}{s_h+s_f+s_d+s_b}
\ms{-\frac\ga2}{h}{s_g+s_e+s_c+s_a}
\ms{\pit}{g}{s_f+s_d+s_b}
\\
\Ms{0}{f}
\ms{-\pit}{e}{s_d+s_b}
\ms{\frac\ga2}{d}{s_c+s_a}
\ms{\frac{\pi-\da-\ba}2}{c}{s_b}
\Ms{0}{b}
\Ms{\frac{-\ba+\da+\pi}2}{a}
\\
\et{B}{C}\et{A}{B}
\et{j}{k}\et{i}{j}\et{h}{i}
\et{g}{h}\et{f}{g}\et{A}f\et{e}{f}\et{d}{e}\et{c}{d}\et{b}{c}\et{a}{b}\et{A}b
}
Figure~\ref{dependgraph} shows the dependency structure of the resulting
standard pattern for $\ctR U$, and one sees it has depth complexity $7$.

\section{The no dependency theorems}
From standardization we can also infer results related dependencies.
We start with a simple observation which is a direct consequence of standardisation.

\LE
Let $\mfr P$ be a pattern implementing some unitary $U$, and suppose $\mfr
P$'s command sequence has measurements only of the $M^x$ and $M^y$ kind,
then $U$ has a standard implementation, 
having only independent measurements, all being of the $M^x$ and $M^y$ kind
(therefore of depth complexity at most 2).
\EL
Write $\mfr P'$ for the standard pattern associated to $\mfr P$.
By equations (\ref{mx}) and (\ref{my}), the $X$-actions can be eliminated
from $\mfr P'$, and then $Z$-actions can be eliminated by using the extended calculus. The final pattern still implements $U$, has no longer any dependent measurements, and has therefore depth complexity at most 2.
\qed

\TH 
Let $U$ be a unitary operator, 
then $U$ is in the Clifford group iff
there exists a pattern $\mfr P$ implementing $U$, 
having measurements only of the $M^x$ and $M^y$ kind. 
\HT
The ``only if'' direction is easy,
since we have seen in the example section, standard patterns for $\ctR X$, $H$ and $P(\pit)$ which had only $M^x$ and $M^y$ measurements. Hence any Clifford operator can be implemented by a combination of these patterns. By the lemma above, we know we can actually choose these patterns to be standard.

For the ``if'' direction, we prove that $U$ belongs to the normaliser of the Pauli group, and hence by definition to the Clifford group. In order to do so we use the standard form of $\mfr P$ written as $\mfr P'= C_{\mfr P'}M_{\mfr P'}E_{\mfr P'}$ which still implements $U$,
and has only $M^x$ and $M^y$ measurements.

Let $i$ be an input qubit, and consider the pattern $\mfr P''={\mfr P}C_i$, where $C_i$ is either $X_i$ or $Z_i$. Clearly $\mfr P''$ implements $UC_i$. 
First, one has:
\AR{
C_{\mfr P'}M_{\mfr P'}E_{\mfr P'} C_i 
&\Rar_{EC}\st& 
C_{\mfr P'}M_{\mfr P'}C' E_{\mfr P'}
}
for some \emph{non-dependent} sequence of corrections $C'$,
which, up to free commutations can be written uniquely
as $C'_OC''$, where $C'_O$ applies on output qubits, and therefore
commutes to $M_{\mfr P'}$, and $C''$ applies on non-output qubits
(which are therefore all measured in $M_{\mfr P'}$).
So, by commuting $C'_O$ both through $M_{\mfr P'}$ and $C_{\mfr P'}$ (up to a global phase), one gets:
\AR{
C_{\mfr P'}M_{\mfr P'}C' E_{\mfr P'}
&\Rar\st& 
C'_OC_{\mfr P'}M_{\mfr P'}C'' E_{\mfr P'}
}
Using equations (\ref{mx}), (\ref{my}), and the extended calculus
to eliminate the remaining $Z$-actions, one gets:
\AR{
M_{\mfr P'}C''
&\Rar_{MC,S}\st& 
SM_{\mfr P'}
}
for some product $S=\prod_{\ens{j\in J}}\ss j1$ of constant shiftings, applying to some subset $J$ of the non-output qubits. So:
\AR{
C'_OC_{\mfr P'}M_{\mfr P'}C'' E_{\mfr P'}
&\Rar\st& 
C'_OC_{\mfr P'}SM_{\mfr P'}E_{\mfr P'}\\
&\Rar\st& 
C'_OC''_OC_{\mfr P'}M_{\mfr P'}E_{\mfr P'}
}
where $C''_O$ is a further constant correction obtained
by shifting $C_{\mfr P'}$ with $S$. This proves that 
$\mfr P''$ also implements $C'_OC''_OU$, and therefore 
$UC_i=C'_OC''_OU$ which completes the proof, since
$C'_OC''_O$ is a non dependent correction. 
\qed

The only if part of this theorem already appears in previous work~\cite[p.18]{mqqcs}.
\smallskip

We can further prove that dependencies are crucial for
the universality of the model. Observe first that if a pattern has
no measurements, and hence no dependencies, then it follows from (D2) that
$V=O$, \ie, all qubits are outputs. Therefore computation steps involve
only $X$, $Z$ and $\ctR Z$, and it is not surprising that they compute a
unitary which is in the Clifford group. The general argument essentially
consists in showing that when there are measurements, but still no
dependencies, then the measurements are playing no part in the result. 
\TH
Let $\mfr P$ be a pattern implementing some unitary $U$, and suppose $\mfr
P$'s command sequence doesn't have any dependencies,  
then $U$ is in the Clifford group.
\HT
Write $\mfr P'$ for the standard pattern associated to $\mfr P$.
Since rewriting is sound, $\mfr P'$ still implements $U$, and 
since rewriting never creates any dependency, it still has no dependencies. 
In particular, the corrections one finds at the
end of $\mfr P'$, call them $C$, bear no dependencies.
Erasing them off $\mfr P'$, results in a pattern $\mfr P''$ which is still
standard, still deterministic, and implementing $U':=C\ad U$. 

Now how does the pattern $\mfr P''$ run on some 
input $\phi$~? First $\phi\otimes\ket{\hskip-.4ex+\ldots+}$ goes by the
entanglement 
phase to some $\psi\in\hil V$, and is then subjected to a sequence of
independent 1-qubit measurements. 
Pick a basis $\mcl B$ spanning the Hilbert
space generated by the non-output qubits $\hil{V\setminus O}$
and associated to this sequence of measurements.

Since $\hil V=\hil O\otimes \hil{V\setminus O}$
and $\hil{V\setminus O}=\oplus_{\phi_b\in B}[\phi_b]$,
where $[\phi_b]$  is the linear subspace generated
by $\phi_b$, by distributivity, $\psi$ uniquely 
decomposes as:
\AR{
\psi=\sum_{\phi_b\in\mcl B} \phi_b\otimes x_b
}
where $\phi_b$ ranges over $\mcl B$,
and $x_b\in\hil O$. Now since $\mfr P''$ is deterministic, there exists an
$x$, and scalars $\la_b$ such that $x_b=\la_b x$. Therefore $\psi$ can be
written $\psi'\otimes x$, for some $\psi'$.  
It follows in particular that the output of the computation will still be
$x$ (up to a scalar), no matter what the actual measurements are.  One can
therefore choose them to be all of the $\M x{}$ kind, 
and by the preceding theorem $U'$ is in the Clifford group, and so is $U=CU'$, since $C$ is a Pauli operator. 
\qed

From this section, we conclude in particular that any universal set of patterns
has to include dependencies (by the preceding theorem), and also needs
to use measurements $M^\al$ where $\al\neq0$ modulo $\pit$ (by the theorem
before). This is indeed the case for the universal set $\mfr\G(\al)$ and $\ctR{\mfr Z}$.

\section{Conclusion}
We presented a calculus for 1-qubit measurement based quantum computing.  We
have seen that pattern combinations allow for a structured proof of
universality, which also results in parsimonious implementations.  We have
shown further that our calculus defines a quadratic-time standardisation
algorithm transforming any pattern to a standard form where entanglement is
done first, then measurements, then local corrections. And finally, we have inferred from this procedure that patterns with no dependencies, or using
only Pauli measurements, may only implement unitaries in the Clifford group.

An obvious question is whether one can extend these ideas to other
measurement based models, perhaps based on different families of
entanglement operators, more general measurements and other types of local
corrections.  This is a matter which we wish to explore further.  For now,
it is already clear that both the notation and the calculus can be extended
to the teleportation model which is based on 2-qubit measurements.  This
actually shows that teleportation models are embeddable in the one-way
model in a very strong sense.  We will return to this particular question
elsewhere.

We also feel that the methods explored here can be stretched further and
made to be relevant to the study of error propagation and error correcting,
but this demands using mixed states, and interpreting patterns as cp-maps.  

Finally, there is also a clear reading of dependencies as classical
communications, while local corrections can be thought of as local quantum
operations in a multipartite scenario. Along this reading, standardisation 
pushes non-local operations to the beginning of a distributed computation, and
it seems the measurement calculus could prove useful 
in the area of quantum protocols.

\section{Appendix}
We prove here that standardisation has indeed the properties quoted in the
body of the paper. First, we need a lemma:
\LE[Termination] 
For all $\mfr P$, there exists finitely many $\mfr P'$ such that $\mfr P\Rar\st\mfr P'$. 
\EL 
Suppose $\mfr P$ has command sequence $A_n\ldots A_1$, and define for $A_i=\et ij$ $d(A_i)=i$, and for
$A_j=\cx us$, $d(A_j)=n-j$. 
Define further: 
\AR{ d(\mfr P)&=&( \sum_{E\in\mfr P}d(E), \sum_{C\in\mfr P}d(C)) 
} 
This measure
decreases lexicographically under rewriting, in other words $\mfr P\Rar\mfr P'$ implies $d(\mfr P)>d(\mfr P')$, where $<$ is the
lexicographic ordering on $\mbb N^2$. Let us inspect all cases. 
First when one applies $EC$, then the first coordinate strictly diminishes
(the second does not always, because of the duplication involved if $C=X$); when $MC$, the second strictly diminishes and the first stays
the same or diminishes; when $EA$, the first strictly diminishes (because we dropped the case when $A$ is itself an $E$), and maybe the
second; when $AC$, the second strictly diminishes, and the first stays the same or diminishes (when $A=E$).

Therefore, all rewritings are finite, and since the system is finitely branching (there are no more than $n$ possible single step rewrites on a given sequence of length $n$), we get the statement of the theorem. 
\qed

It is not to difficult to strengthen the result above, by showing that the longest possible rewriting of $\mfr P$ is quadratic in $n$, where
$n$ is the length of $\mfr P$'s command sequence.

Say $\mfr P$ is \emph{standard} if for no $\mfr P'$, $\mfr P\Rar\mfr P'$.

\PRO[Standardisation] 
For all $\mfr P$, there exists a unique standard $\mfr P'$, such that $\mfr P\Rar\st\mfr P'$, and $\mfr P'$ satisfies the (EMC) condition. 
\ORP 
Since the rewriting system
is terminating, confluence follows from local confluence 
(meaning whenever two rewritings can be applied, one can
rewrite further both transforms to a same third expression). 
Then, uniqueness of the standard form is an easy consequence (actually, for
terminating rewriting systems, unicity of standard forms and confluence are equivalent). Looking for critical pairs, that is occurrences of three
successive commands where two rules can be applied simultaneously, 
one finds that there are only two types: $\et ijM_kC_k$ with $i$, $j$
and $k$ all distinct, and $\et ijM_kC_{l}$ with $k$ and $l$ distinct. 
In both cases local confluence is easily verified. 

Suppose now ${\mfr P'}$ does not satisfy (EMC). Then, either there is a pattern $EA$ with $A$ not of type $E$, or there is a
pattern $AC$ with $A$ not of type $C$. In the former case, $E$ and $A$ must operate on overlapping qubits, else one may apply a free
commutation rule, and $A$ may not be a $C$ since in this case one may apply an $EC$ rewrite. The only remaining case in when $A$ is of type
$M$, overlapping $E$'s qubits, but this is what condition (D1) forbids, 
and since (D1) is preserved under rewriting, this contradicts the
assumption. The latter case is even simpler. 
\qed

\subsection{Discussion}
This is what we wanted, namely we have shown that under rewriting any 
pattern can be put in (EMC) form. We actually proved a bit more,
namely that the standard form obtained is unique. 

However, one has to
be a bit careful about the significance of this additional piece
of information. Note first that unicity is obtained because we dropped the $CC$ free commutations, and all $EE$ commutations, thus having a very rigid notion of
command sequence. One cannot put them back as rewriting rules, since they obviously ruin termination and uniqueness of standard forms.

A reasonable thing to do, would be to take this set of equations as generating an  equivalence relation on command sequences, call it $\equiv$,
and hope to strengthen the results obtained so far, by proving that all reachable standard forms are equivalent.

But this is too naive a strategy, since 
$\et12\Cx1\Cx2\equiv\et12\Cx2\Cx1$, and: 
\AR{ \et12\cx1s\cx2t
&\Rar\st&\cx1s\cz2s\cx2t\cz1t\et12\\
&\equiv&\cx1s\cz1t\cz2s\cx2t\et12 
} 
obtaining an expression which is not symmetric in $1$ and $2$. To conclude, one has to extend $\equiv$ to include the additional equivalence $\cx1s\cz1t\equiv\cz1t\cx1s$, which fortunately is sound since these two operators are equal up to a global phase. 
We conjecture that this enriched equivalence is preserved.

\end{document}